\documentclass[journal]{IEEEtran}
\ifCLASSINFOpdf
\else
\fi
\usepackage{amsfonts}
\usepackage{dsfont}
\usepackage{marvosym}

\usepackage{amssymb}
\usepackage[mathscr]{euscript}
\usepackage{bm}

\usepackage{graphicx}
\usepackage{subfig}

\usepackage[noadjust]{cite}

\usepackage{amsmath}
\usepackage{amsthm}

\usepackage[ruled,vlined]{algorithm2e}
\usepackage{color}
\usepackage[normalem]{ulem}
\usepackage{comment}
\usepackage{etoolbox}
\usepackage{mathtools}
\makeatletter
\patchcmd{\@algocf@start}
  {-1.5em}
  {0pt}
  {}{}
\makeatother

\usepackage{titlesec}
\titlespacing{\section}{0pt}{*1.2}{*0.5}
\titlespacing{\subsection}{0pt}{*0.5}{*0.5}

\usepackage[table]{xcolor}
\usepackage{multirow}
\usepackage{ctable}
\usepackage{balance}
\def\X{\mathcal{X}}

\def\M{\mathcal{M}}

\def\Om{\mathcal{O}}
\def\e{\varepsilon}
\def\Prob{\mathrm{Pr}}

\def\Lap{\mathrm{Lap}}

\newcolumntype{I}{!{\vrule width 1pt}}
\newlength\savedwidth

\newcommand\so{\bgroup\markoverwith{\textcolor{red}{\rule[0.5ex]{2pt}{0.4pt}}}\ULon}

\newtheorem{definition}{Definition}

\newtheorem{remark}{Remark}

\begin{document}

\title{Realistic Differentially-Private Transmission Power Flow Data Release}

\author{David Smith, Frederik Geth,  Elliott Vercoe, Andrew Feutrill, Ming Ding, Jonathan Chan, James Foster and Thierry Rakotoarivelo
\thanks{D. Smith, E. Vercoe, A. Feutrill, M. Ding, J. Chan and T. Rakotoarivelo are with Data61, CSIRO (Commonwealth Scientific and Industrial Research Organisation),  Australia (e-mail: \{David.Smith, Elliott.Vercoe, Andrew.Feutrill, Ming.Ding, Jonathan.Chan, Thierry.Rakotoarivelo\}@data61.csiro.au).}
\thanks{F. Geth and J. Foster are with Energy, CSIRO, Australia (email:\{Frederik.Geth, James.Foster\}@csiro.au).}}


\maketitle
\begin{abstract}
For the modeling, design and planning of future energy transmission networks, it
is vital for stakeholders to access faithful and useful power
flow data, while provably maintaining the privacy of business
confidentiality of service providers. This critical challenge has recently been
somewhat addressed in [1]. This paper significantly extends this existing work.
First, we reduce the potential leakage information by proposing a fundamentally
different post-processing method, using public information of grid losses rather
than power dispatch, which achieve a higher level of privacy protection. Second,
we protect more sensitive parameters, i.e., branch shunt susceptance in addition to series impedance (complete pi-model). This protects power flow data for the transmission high-voltage
networks, using differentially private transformations that maintain the optimal
power flow consistent with, and faithful to, expected model behaviour. Third, we
tested our approach at a larger scale than previous work, using the PGLib-OPF
test cases [10]. This resulted in the successful obfuscation of up to a 4700-bus
system, which can be successfully solved with faithfulness of parameters and good
utility to data analysts. Our approach addresses a more feasible and realistic
scenario, and provides higher than state-of-the-art privacy guarantees, while
maintaining solvability, fidelity and feasibility of the system.
\end{abstract}

\IEEEpeerreviewmaketitle
\section{Introduction}


Open data access in smart grids, 
including network parameters, voltage measurements, energy consumption, etc., 
is envisaged to enable the development of powerful decision-support tools for transmission network service providers and distribution system operators.
However, across the network from transmission to distribution such an open data
approach raises business confidentiality risks for network providers and
individual privacy risks for grid customers e.g., the value parameters of lines
and generators may reveal how transmission operators operate their networks; or
at the distribution-side exposing an individual's energy use patterns may reveal
household habits or composition. 
Furthermore, transmission-side data, such as network topology or generation patterns,
could be exploited by an attacker to inflict targeted damages
on the network infrastructure or could be exploited as commercially sensitive
information to gain financial benefits.


In this context, a recent work by Fioretto et al.~\cite{fioretto2020} proposed a
framework based on Differential Privacy to allow the release of power network
data in a provable private and confidential manner. Their framework is designed
around pre and post-processing methods, and the produced obfuscated data was
evaluated in the context of the complex Optimal Power Flow (OPF) problem.
However their proposed post-processing steps~\cite{fioretto2020} rely on
appropriate adjustments to their private solution to provide fidelity, which
requires the dispatch information obtained from the original data. This reliance
on inputs from the original data likely results in privacy leakage, i.e., the
post-processing is implicitly based on the non-obfuscated data~\cite{dwork2014}.


Importantly, as noted in~\cite{fioretto2020}, applying differential privacy faces significant challenges when the resulting
privacy-preserving datasets are used as inputs to complex
physics simulation problems, e.g.,  power flow in transmission grids
problems. Indeed, the privacy-preserving dataset may lose
the fidelity and realism of the original data and may even
not admit feasible solutions for the physics of
interest.
While such problem is addressed in~\cite{fioretto2020}, this paper significantly
extends on that previous solution with a more realistic modeling, a reduced privacy
leakage in the post-processing stage (after applying differential privacy), and
an implementation evaluation at a much larger scale.


In this paper, 
we seek to design a system that can publish transformed power flow information, 
faithful to the expected system behaviour, 
with sufficient accuracy, 
and yet does not compromise the  sensitivity with respect to protecting network impedance data. 

To this end we propose a differential privacy mechanism~\cite{dwork2006,dwork2014},
which makes several practical and important changes to the algorithm from
Fioretto et al.~\cite{fioretto2020}. These changes provide more utility,
represent the optimal power flow more accurately, and do not potentially leak
information. 
Hence:


\begin{enumerate}
\item For the transmission network problem, particularly concerned with preserving generator set-points, we achieve the preservation of power flow faithfulness, with better differential privacy modeling and more representative transmission grid modeling as follows:
\begin{enumerate}
\item For the parameter of admittance, 
we choose to first privately transform series reactance (instead of resistance) as it contains more information than resistance in transmission networks, due to their strongly inductive behavior~\cite{wood2013,dhople2015}. It is well-known that in meshed transmission systems, the `dc' approximation to the power flow equations, which only uses series reactance in the representation of the physics and neglects resistance, provides power flow results that strongly correlate with the underlying physics~\cite{purchala2005}. 
Therefore, we scale resistance according to ratio to reactance in the original data. 
This is the converse case to ~\cite{fioretto2020} and provides better utility and privacy.
\item We change the post-processing after the privacy-preserved transformation.
Indeed, \cite{fioretto2020} uses the power flow optimal dispatch cost objective,
which was derived from the original data, leading to information leakage. We use
instead publicly available grid losses information~\cite{nationalgrideso, aemo},
which does not leak information from the original data.
\item We add branch shunt admittance to model (branch shunt conductance is typically negligible, and is set to 0 in numerous data sets).
\item The bounds in the main algorithm are updated and reversed from~\cite{fioretto2020} as series susceptance is negative ($b=-1/x,\; x>0$, in real world).
\item We apply appropriate bounding on conductance and susceptance to remain feasible.
\end{enumerate}
\item Implementation of nonlinear optimization (up to cubic complex-valued polynomials), 
that accurately and feasibly solved on higher scale problems with the
differentially private transformed data.
\item Implementation of a differentially private method, using Laplace distributed random perturbations,
with the appropriate sensitivity parameter, constraint bounds, and a typical privacy budget $\e=1$.
\item The problem is feasible and solvable for larger data sets with the proposed changes.
\item We implemented our solution in a software package that builds on top of the
heavily-tested PowerModels.jl package~\cite{powermodels}. Our implementation, the PowerModelsPrivacyPreserving.jl, is publicly available for
peer-review and further usage\footnote{https://github.com/frederikgeth/PowerModelsPrivacyPreserving.jl}.
It contains additional unit tests for the required extra functionality, and
was validated using the PG lib benchmark library~\cite{babaeine2019}.
It managed to obfuscate up to 4700 bus system (i.e. more than a factor of 10 larger than that of ~\cite{fioretto2020}), and output faithful power flow results back to matpower `.m' files.
\end{enumerate}

Thus, here we provide a provable privacy-preserving, realistic and practical, optimal power flow-based approach of preserving feasibility for transmission networks, that successfully operates up to a large-scale in terms of number of buses, and could be feasibly implemented to protect network impedance data. This is further explained and detailed in the remainder of this paper, which is outlined as follows. In Section II we briefly survey related work. In Section III we provide some preliminaries, particularly with respect to differential privacy. In Section IV we provide the system model, encapsulating optimal power flow model justification, threat model and the mechanism, including the general realistic optimal power flow model and the algorithm for privacy-preserved minimum reactive power losses in optimal power flow. In Section V we provide some comprehensive experimental evaluations using the well known PGLib-OPF test cases up to a very large number of buses, with high-utility privacy preserved outputs.
We further highlight some results for different numbers of buses and levels of indistinguishability. Finally in Section VI we provide some concluding remarks and propose areas for future work. 
\begin{table*}[!t]
\renewcommand{\arraystretch}{1.2}
\vspace{-5mm}
\caption{Power Network Nomenclature}
\label{t:notations}
\vspace{-2mm}
\centering
\setlength\tabcolsep{2pt}
{\begin{tabular}{l l l l l l l l l l l l}
\hline
$\mathbf{b}$ & Series susceptances of $\mathcal{N}$ & & & & $N$ & The set of nodes in the network & & & & $\mathbf{\dot x}$ & Post-processed version of $\mathbf{x}$\\
$\mathbf{b^{sh}}$ & Shunt susceptances of $\mathcal{N}$ & & & & $\mathcal{N}$ & A network description & & & & $\mathbf{\tilde x}$ & Privacy-preserving version of $\mathbf{x}$ \\
$c_0, c_1, c_2$ & Generation cost coefficients & & & & $\mathbf{r}$ & The vector of ratios $\{{\textsl{g}_{l}}/{b_{l}}\}_{(lij) \in E}$ & & & & $\mathbf Y$ & Series admittances of $\mathcal{N}$\\
$E$ & The set of {\it from} edges in the network & & & & $\mathfrak{R}(.)$ & Real component of a complex number & & & & $\mathbf{Y^{sh}}$ & Shunt admittances of $\mathcal{N}$ \\
$E^R$ & The set of {\it to} edges in the network & & & & $s^u$ & Line apparent power thermal limit & & & & $\alpha$ & Indistinguishability value\\
$E(v)$ & The subset of lines at voltage level $v$ & & & & $S$ & AC power & & & & $\beta$ & Objective faithfulness factor\\
$\mathbf{g}$ & Series conductances of $\mathcal{N}$ & & & & $S^d$ & AC power demand & & & & $\e$ & Privacy budget\\
$\mathbf{g^{sh}}$ & Shunt conductances of $\mathcal{N}$ & & & & $S^\textsl{g}$ & AC power generation & & & & $\theta_{ij}$ & Phase angle difference (i.e., $\theta_i - \theta_j)$\\
$\mathcal{L}^*(N)$ & Optimal grid loss of $\mathcal{N}$ & & & & $V$ & AC voltage & & & & $\theta^\Delta$ & Phase angle difference limits\\
$m$ & $|N|$ & & & & $\mathit{VL} (\mathcal N)$ & The set of voltage levels in $\mathcal{N}$ & & & & $\lambda$ & Optimization constraint scale factor\\
$n$ & $|E|$ & & & & $x^l, x^u$ & Lower and upper bounds of $x$ & & & & $\tilde{\mu}_\mathbf{x}$ & The noisy mean value of the vector $\mathbf{x}$\\
$n_v$ & $|E(v)|$ & & & &  $\mathbf{x}^*$ & Complex conjugate of $\mathbf{x}$ & & & & $\angle$ & Angle of a complex number\\
\hline
\vspace{-4mm}
\end{tabular}}
\end{table*}

\section{Related Work}
With the increase in research interest in power systems modeling, a number of datasets and tools have been developed to address the shortfall in available data and enable scientists and engineers to apply techniques to power network optimization problems.

For example, Coffrin et al. conducted a survey of publicly available test case
data for use in the context of AC Optimal Power Flow (AC-OPF) problems~\cite{coffrin2014n}.
They studied the test cases from the Matpower tools as well as several IEEE test
cases, and proposed NESTA, a novel data-driven approach to improve on them. However
they acknowledged that the NESTA-generated synthetic test cases are still far from
detailed real-world data.
This data set ended up being the starting point for PG-lib ~\cite{babaeine2019}, a repository that also collects benchmarks for unit commitment and OPF problems with high-voltage DC components. 
The Julia programming language has emerged as a de-facto tool in the power system
modeling space. Specifically, the  PowerModels.jl package~\cite{powermodels} provides many
functions to solve network simulation and optimization problems for transmission networks,
under a variety of constraints.
In this paper, we extend PowerModels.jl as part of  a new package PowerModelsPrivacyPreserving.jl to provide power systems data release with differential privacy guarantees.

Several approaches exist to solve the OPF problem's constrained optimizations,
including relaxation methods to transform them into convex ones. Examples of
these convex relaxations of the OPF constraints were based on Semi-Definite 
Programming~\cite{bai2008}, Quadratic Convex relaxation~\cite{hijazi2017c} and second-order cone programming~\cite{jabr2006r}. Analysis from Coffrin~\cite{coffrin2015qc} showed relationships between these classes of relaxations, and both the Semi-definite Programming and Quadratic Convex are stronger relaxations of Second-Order Cone programming, however Semi-Definite programming and Quadratic Convex are not equivalent. These relaxations have since been combined with differential privacy perturbations to find near optimal solutions to the OPF problem~\cite{fioretto2018, fioretto2020}.

To maintain high fidelity when using differential privacy, a bilevel optimization approach is described in \cite{mak2020} to optimally redistribute the noise by a randomized mechanism to meet both indistinguishability and fidelity requirements for OPF customer load. When variables to be optimized are a function of the noise instead of a simple noise addition, Dvorkin et al. \cite{dvorkin2020b} exploit a privacy framework to solve constrained convex optimization problems that make these variables differentially private with strong feasibility guarantees. To avoid the complexity of post-processing to restore fidelity, the same researchers have recently introduced a privacy-preserving mechanism which parameterizes OPF variables as affine functions of the noise such that their correlations with grid loads can be weakened to better protect the privacy of load consumption from voltage and power flow measurements \cite{dvorkin2020a}. Private computations have also been studied in \cite{zhou2019d} to release aggregated OPF statistics but in the context of direct current (DC) OPF. Bienstock and Shukla \cite{bienstock2019} propose a number of convex formulations of DC-OPF that trade variance in power-flow-related parameters for operational cost. Recent work demonstrates convergence for control of optimal power flow according to a differentially private projected subgradient method~\cite{ryu2021}.

As a different line of work, Karapetyan et al. \cite{karapetyan2017a} study the trade-off between privacy and fidelity in the context of micro-grid based on a privacy-preserving demand response optimization problem. Zhao et al. \cite{zhao2014} investigate the charging/discharging of household batteries in the differential privacy context to address privacy concern of smart meters and Eibl et al. \cite{eibl2017d} study the use case of privacy preserving electric load forecasting. Liao et al. \cite{liao2019} introduce a peak-time load balancing mechanism based on distributed differential privacy. A Stackelberg game energy-trading scheme is studied in \cite{smith2020}, which uses differential privacy to protect household payment data and net energy consumption while maintaining mutual benefits among the households, the community battery operators and the smart grid. Han et al. \cite{han2016} study an $\e$-differentially private distributed constrained optimization in the case of electric vehicle charging.



\section{Preliminaries}

This section introduces the notion of differential privacy that is adopted in
this paper to protect private and confidential information in power flow data.  

\subsection{Differential Privacy}

Applying differential privacy to data is designed to protect an individual from having sensitive information disclosed from a public dataset~\cite{dwork2014}. 
In this paper, 
we propose a differential privacy mechanism for transmission network service providers to publicly share transmission data without additional substantive privacy risk implications.

To that end, we require the notion of a randomised algorithm, 
which is an algorithm that takes some random input for its execution.
More concretely, 
it is a mapping from a domain to a probability space, 
composed of coin flips of the algorithm, 
perturbing the outputs.

\begin{definition}
We say a randomized algorithm, $\M $, 
guarantees $\e$-differential privacy, 
which is known as differential privacy, 
if for any two datasets $\X^{(1)}$ and $\X^{(2)}$, 
differing on at most one coordinate (or record), 
the following property is satisfied

\begin{align}
\frac{\Prob{[\M(\X^{(1)}) \in \Om]}}{\Prob{[\M(\X^{(2)}) \in \Om]}} \leq \exp(\e).
\end{align}
We refer to $\e > 0$ as the \emph{privacy budget}, 
which is a parameter that controls the level of privacy, 
with the smaller the value of $\e$, 
the stronger the privacy protection level. 
\end{definition}

\begin{remark}
If $\Om$ is a countable set, then we have the inequality for each $x \in \Om$,

\begin{equation}
\Prob[\M(\X^{(1)})=x] \leq \exp(\e) \cdot \Prob[\M(\X^{(2)})=x]
\end{equation}
where the probability space is defined as coin flips of $\M$~\cite{dwork2014}.\\
\end{remark}

We denote a random variable drawn from a Laplace (symmetric  exponential) distribution with mean 0 and scale $\lambda$ as $Y \sim \Lap(\lambda)$, 
which has the following probability density function

\begin{equation*}
    f(x|\lambda) = \frac{1}{2\lambda} \exp{\left(-\frac{|x|}{\lambda}\right)}.
\end{equation*}
For a linear query, 
it has been widely demonstrated that adding noise from a zero-mean Laplace distribution with scale $\lambda=\Delta f / \e$, ``the Laplace mechanism", preserves $\e$-differential privacy for real vector valued queries~\cite{dwork2014}, where

\begin{equation*}
    \Delta f = \max_{\X^{(1)} \sim \X^{(2)}} ||f(\X^{(1)}) - f(\X^{(2)})||_1,
\end{equation*}
 indicates the global sensitivity of the query type, 
 which captures the worst case of the amount of change when removing a particular coordinate (or record) from a dataset. 
 The sensitivity therefore is related to how much noise is required to protect the privacy of an individual coordinate in the dataset, 
 as it forms a reference scale for the noise.
 Here we refer to this as the level of indistinguishability, $\alpha=\Delta f$, 
 such that queries on datasets we are interrogating (differing on a single value by at most $\alpha$) will give similar results.
 
\begin{remark}
Some useful properties~\cite{dwork2014} make differential privacy ideal for
building provable privacy-preserving mechanisms.
 These include: 
 \begin{enumerate}
\item Protection against arbitrary risks, 
 \emph{i.e.}, no additional substantive risk is created by sharing differentially private data; 
\item  Quantification of privacy loss; 
 \item Composition of privacy budgets, which enables algorithms to be split up and analyzed step by step;
 \item Immunity to post-processing, i.e. any transformations to the output
 of a differentially private mechanism will produce results that remain
 differentially private.
\end{enumerate}
 \end{remark}

\subsection{Motivation for Threat Mitigation}

To facilitate the management of power system networks, 
there is a critical need 
for open-access to high-fidelity and large-scale power system models. 
The Data Repository for Power System Models project \cite{arpa-e} of ARPA-E in the US, 
and the data released by the AEMO \cite{nemweb} and Geoscience Australia~\cite{ga-substation, ga-line} are examples of such effort. 
However, 
the release of these datasets with a high degree of fidelity may raise concerns of commercial confidentiality and household/customer privacy. 
Moreover, 
due to the physical vulnerabilities~\cite{bienstock2010nk, salmeron2009worst} and cyber security issues~\cite{liu2011false, yuan2011modeling} of the power grid, 
an adversary could exploit these system weaknesses using the published data and cause significant harms. 

In the past two decades, 
there have been many well-organised cyber attacks on the power system, 
such as a hack into the US power grid in 2009~\cite{gorman2009e} that left behind a trail of malicious software in their computer system; 
and the malware attack of the Ukraine's power grid in 2015~\cite{zetter2016i}, 
in which circuit breakers were deliberately turned off, 
and computer hard disks and firmware on critical devices were overwritten. 
We envisage that similar intrusions to power systems will cause even more devastating consequences, 
if the attacker is able to thoroughly exploit the weaknesses of his targeted power grid network using its realistic system model.


\section{System Model}

This section describes the system model underlying our proposed approach.
Table~\ref{t:notations} defines the related notations, which are used throughout this paper.

\subsection{Power Flow Model Choice}

The power flow equations are the core of the large majority of computations for designing and operating electric grids.
This system of multivariate nonlinear equations model the steady-state relationship between complex voltage phasors and power injections in such a power system~\cite{mehta2016}.

Power flow models are then chosen on the basis of both capturing the most relevant aspects of the physics  while remaining computationally tractable. 
The problem is phrased as a root-finding problem, commonly approached through Newton's method, to provide voltages and current throughout the network, given load and generator setpoints as well as Kirchoff's laws.
The choice of model in this work is guided by a focus on line parameters. Therefore an accurate representation of line parameters and the explicit dependence on complex voltages is important for fidelity to the physics, especially to capture sensitivities to line parameters.

\subsection{Threat Model}
Line parameters are extremely sensitive data as they reveal important operational
information that attackers can exploit to inflict targeted damage to network
infrastructure~\cite{fioretto2020}.
As in~\cite{fioretto2020}, we assume that the ratios between conductance and
susceptance of a line can be retrieved from the manufacturer information material.
Our objective is then to protect a network description $\mathcal{N}=\langle N, E, \mathbf{S}, \mathbf{\dot Y}, \bm{\theta^\Delta}, \mathbf{s, v} \rangle$ from an attacker, by providing
obfuscated line admittance values $\mathbf{\dot Y}$ within a given level of
indistinguishability $\alpha$. Importantly we also consider shunt reactance as
a line impedance parameter to be protected, which is not addressed in~\cite{fioretto2020}.

Our approach first obfuscates power flow data sets, and then uses OPF as a tool to
restore feasibility. We further protect any given $k$ number of lines of active
flows given the knowledge of network description $\mathcal{N}$ without the complete
set of line admittance values $\mathbf{Y}$ (all line admittance values separate
to any $k$ target lines, for $k\geq 1$). This encapsulates the plausible
deniability guarantee of differential privacy~\cite{dwork2014}.
In contrast to~\cite{fioretto2020}, our method does not require the optimal dispatch
cost to be publicly available, and instead uses readily publicly available grid
losses~\cite{nationalgrideso, aemo}.

\subsection{The Mechanism}

Similar to~\cite{fioretto2020}, we apply differential privacy with a practical
level of sensitivity and distinguishability whereby the model maintains fidelity
to the ground-truth optimal power flow at the transmission side of the power grid
network. However in addition to the previously discussed differences in the objectives
and assumptions, our proposed mechanism also differs from~\cite{fioretto2020} on
three key modifications and extensions, as detailed below.

First, in~\cite{fioretto2020} the values of conductance were perturbed with differential privacy, then the series reactance scaled proportionally to the resistance. However, here we apply the converse, perturbing the reactance $\mathbf{r}$, then accordingly scaling the resistance $\mathbf{x}$ --- as reactance plays a much stronger role in the coupling between active power flows and voltage angle differentials over lines compared to resistance~\cite{wood2013}.

Second, in~\cite{fioretto2020} an optimal dispatch cost objective was introduced into the objective function after applying differential privacy to restore feasibility after adding noise to network impedance. Although differential privacy is immune to post-processing as noted in ~\cite{fioretto2020}, using the optimal dispatch cost objective may leak sensitive information, as it may give insight into the system solution by allowing an ``inverse optimization" problem to be more easily solved, reconstructing the state of the original problem. Rather we apply transmission-level grid losses $\mathcal{L}$ as these are a more technically-focused aggregated metric of the system, so that there are many states/configurations that may produce the approximately the same grid loss value.

Finally, as opposed to ~\cite{fioretto2020} we also obfuscate the branch shunt susceptance of the pi-model, $\mathbf{b^{sh}}$ (we assume shunt conductance is zero). Shunt susceptance becomes particularly prominent for cables and/or longer overhead lines, and can have significant influence on the flow of reactive power in the network.

Together, these three key differences lead to a model of the AC Optimal Power
Flow problem, which is more realistic than in~\cite{fioretto2020} and will
be detailed further next.

\begin{definition}
We are using a generalization of the canonical AC optimal power flow (OPF) problem. The formulation is summarized in Model 1, as a function of complex power variables $\mathbf{S}$ and voltages $\mathbf{V}$, at network edges $E$.
It can be observed from Model 1, (9) and (10) that a cubic polynomial is being implemented. The mechanism for privacy-preserved minimum reactive power losses in this AC-OPF, faithful to suitable power flow constraints $(s_1)-(s_6)$, is described in Algorithm 1, with description of OPF by output network description {$\mathcal{\dot N} = \langle N, E, \mathbf{S}, \mathbf{\dot Y}, \bm{\theta^\Delta}, \mathbf{s, v} \rangle$} consisting of the set of nodes $N$, the from edges $E$, transformed AC power $\mathbf{S}$ and node admittance $\mathbf{\dot Y}$, phase angle difference limits \bm{$\theta^\Delta$}, line apparent thermal limits $\mathbf{s}$ and voltage levels $\mathbf{v}$. Equations (\ref{eq:id1}) to (\ref{eq:id15}) describe the perturbation of shunt and series susceptances, and shunt conductances, as well as the $\mu$ parameter, used in ($s_4$) to ($s_6$), to solve the model as scaling constraints.
\end{definition}

\begin{remark}
We design the model to have the following properties:
\begin{enumerate}
    \item \emph{Line obfuscation}: The line \emph{susceptances} $\mathbf{\tilde b}$
    of $\mathcal{\tilde N}$, the network description satisfy $\e$-differential privacy under $\alpha$-indistinguishability
    \item \emph{Consistency}: $\mathcal{\tilde N}$ have feasible solutions to the OPF Constraints Model 1, (4) to (10)
    \item \emph{Objective Faithfulness}: $\mathcal{\tilde N}$ is faithful to the value of the objective function (grid losses) up to a factor $\beta$, i.e., $|\frac{\mathcal{O(N)-O(\tilde N)}}{\mathcal{O(N)}}| \leq \beta$
\end{enumerate}
\end{remark}

\LinesNumberedHidden

\begin{algorithm}[t!]
\SetAlgorithmName{Model}{model}{List of models}
\caption{A realistic model for AC Optimal Power Flow problem}\label{Model1}
\textbf{variables:} $S^\textsl{g}_i, V_i$ \hspace{4em} $\forall i \in N$\\[3pt]
\hspace{4.2em} $Y^{sh}_{lij} = j{{b}}^{sh}_{lij} , Y^*_{l} = {\textsl{g}}_{l} + j{\textsl{b}}_{l},$ $ S_{lij}$ \\ \hspace{11.3em} $\forall (l,i, j) \in E \cup E^R$\\[3pt]
\textbf{minimize:} 
\begin{equation}
\sum_{i \in N} \mathbf{c}_{2i} \left( \mathfrak{R}(S^\textsl{g}_i) \right)^2 +  \mathbf{c}_{1i} \mathfrak{R}(S^\textsl{g}_i) + \mathbf{c}_{0i}
\end{equation}\\
\textbf{subject to:}\\
\Indm
\begin{equation}
\angle V_s = 0
\end{equation}\\
\begin{equation}
\bm{v^l}_i \leq |V_i| \leq \bm{v^u}_i \quad \forall i \in N
\end{equation}\\
\begin{equation}
\bm{-\theta^\Delta}_{ij} \leq \angle \left(V_i V^*_j \right) \leq \bm{\theta^\Delta}_{ij} \quad \forall (l,i, j) \in E
\end{equation}\\
\begin{equation}
\bm{S^{\mathbf gl}}_i \leq S^\textsl{g}_i \leq \bm{S^{\mathbf gu}}_i \quad \forall i \in N
\end{equation}\\
\begin{equation}
|S_{lij}| \leq \bm{s^{u}}_{lij} \quad \forall  (l,i, j) \in E \cup E^R
\end{equation}\\[3pt]
\begin{equation}
\bm{S^{\mathbf g}}_i - \bm{S^{\mathbf d}}_i - \bm{Y^{sh}}_i |V_i|^2 = \sum_{\mathclap{(l,i, j) \in  E \cup E^R}} S_{lij} \quad \forall i \in N
\end{equation}\\[5pt]
\begin{equation}
S_{lij} = (Y^{sh}_{lij})^* |V_i|^2 - Y^*_{l} V_i V^*_j \quad \forall  (l,i, j) \in E \cup E^R
\end{equation}\\
\end{algorithm}

\begin{algorithm}[t!]
\setcounter{algocf}{0}
\caption{Mechanism for privacy-preserved minimum reactive power losses in the the AC-Optimal Power Flow (OPF)}\label{Algo1}
\SetKwInOut{Input}{input}
\SetKwInOut{Output}{output}
\Input{$\mathcal{\langle N, L^*, \e,\alpha,\beta \rangle}$}
\ShowLn
$\mathbf{\tilde{b}} \gets \mathbf{b}+{\mathrm{Lap}\left(\dfrac{3\alpha}{\e}\right)}$\\
\ShowLn
$\mathbf{\tilde{g}} \gets {\frac 1 {\mathbf{r}}} \cdot \mathbf{\tilde{b}}$\\
\ShowLn
$\mathbf{\tilde{b}^{sh}} \gets \mathbf{{b}^{sh}}+{\mathrm{Lap}\left(\dfrac{3\alpha}{\e}\right)}$\\
\ShowLn
\ShowLn
\ForEach{$v \in \mathit{VL} (\mathcal N) $}{%
\ShowLn
  $\tilde{\mu}^v_\mathbf{x} \gets \frac 1 {n_v}\sum_{(lij) \in E(v)} x_{lij} + \mathrm{Lap}\left(\dfrac{3\alpha}{{n_v}\e}\right)$ \hspace{7em} (for $\mathbf{x = g, b,b^{sh}}$)\\
}
\ShowLn
Solve the following model:\\
\Indp
\textbf{variables:} $S^\textsl{g}_i, V_i$ \hspace{4em} $\forall i \in N$\\[3pt]
\hspace{4.2em} $\dot Y^{sh}_{lij} = j\dot{{b}}^{sh}_{lij} , \dot Y^*_{l} = \dot{\textsl{g}}_{l} + j\dot{\textsl{b}}_{l},$ $ S_{lij}$ \\ \hspace{11.3em} $\forall (l,i, j) \in E \cup E^R$\\[3pt]
\textbf{minimize:} $\lVert \mathbf{\dot{g}-\tilde{g}} \rVert^2_2 + \lVert \mathbf{\dot{b}-\tilde{b}} \rVert^2_2$ \hspace{6.1em} $(s_1)$\\[3pt]
\textbf{subject to: } (4)--(10)\\[4pt]
\Indp 
\Indp
$\frac{|\sum_{i \in N}\bm{loss} \left( S^\textsl{g}_i \right)- \mathcal{L^*}|} {\mathcal{L^*}} \leq \beta$ \hspace{6.8em} $(s_2)$\\[3pt]
\Indm
\Indm
$\forall  (l,i, j) \in E \cup E^R:$ \\
\Indp 
\Indp
$S_{lij} =  (\dot Y^{sh}_{lij})^* |V_i|^2 - \dot Y^*_{l} V_i V^*_j$ \hspace{4.3em} $(s_3)$\\[3pt]
\Indm
\Indm
$\forall  (l,i, j) \in E(v) \cup E^R(v), \forall v \in \mathit{VL} (\mathcal N):$ \\[3pt]
\Indp 
\Indp
$\frac{1}{\lambda} \mu^v_\mathbf{g} \leq \dot{\textsl{g}}_{l} \leq \lambda \mu^v_\mathbf{g}$ \hspace{9.7em} $(s_4)$\\[3pt]
${\lambda} \mu^v_\mathbf{b} \leq \dot{b}_{l} \leq \frac{1}{\lambda} \mu^v_\mathbf{b}$ \hspace{9.7em} $(s_5)$\\[3pt]
$\frac{1}{\lambda} \mu^v_\mathbf{b^{sh}} \leq \dot{{b}}^{sh}_{lij} \leq \lambda \mu^v_\mathbf{b^{sh}}$ \hspace{7.5em} $(s_6)$\\[6pt]
\Indm
\Indm
\Indm
\Output{$\mathcal{\dot N} = \langle N, E, \mathbf{S}, \mathbf{\dot Y}, \bm{\theta^\Delta}, \mathbf{s, v} \rangle$}
\end{algorithm}






Then the following equations describe the perturbations of the line properties.

\begin{equation}
\mathbf{\tilde{b}}=\mathbf{b}+\mathrm{Lap}\left(\dfrac{3\alpha}{\e}\right) ,\quad \mathbf{\tilde{g}}=  {\frac 1 {\mathbf{r}}} \cdot \mathbf{\tilde{b}} \label{eq:id1}
\end{equation}\;

\begin{equation}
\mathbf{\tilde{b}^{sh}}=\mathbf{b^{sh}}+\mathrm{Lap}\left(\dfrac{3\alpha}{\e}\right),\quad \mathbf{\tilde{g}^{sh}}= \mathbf{g^{sh}}= 0  
\label{eq:id2}
\end{equation}\;

\begin{equation}
\tilde{\mu}^v_\mathbf{g}=\left(\frac 1 {n_v}\sum_{(lij) \in E(v)} \textsl{g}_{l}\right) + \mathrm{Lap}\left(\dfrac{3\alpha}{{n_v}\e}\right) \label{eq:id13}
\end{equation}\;

\begin{equation}
\tilde{\mu}^v_\mathbf{b}=\left(\frac 1 {n_v}\sum_{(lij) \in E(v)} \textsl{b}_{l}\right) + \mathrm{Lap}\left(\dfrac{3\alpha}{{n_v}\e}\right) \label{eq:id14}
\end{equation}\;

\begin{equation}
\tilde{\mu}^v_\mathbf{b^{sh}}=\left(\frac 1 {n_v}\sum_{(lij) \in E(v)} \textsl{b}^{sh}_{lij}\right) + \mathrm{Lap}\left(\dfrac{3\alpha}{{n_v}\e}\right)
\label{eq:id15}
\end{equation}

\section{Experimental Evaluations}
As a valuable outcome of our study, 
we implemented the ``PowerModelsPrivacyPreserving.jl" (PMPP)${}^1$, 
as an extension package of PowerModels.jl~\cite{powermodels}, to obfuscate values in the sensitive power flow datasets to protect commercial confidentiality.
This section presents the results of evaluating PMPP on collections of well-known power network test cases from the PGLib-OPF~\cite{babaeine2019} repository. 
It validates our privacy-preserving obfuscation mechanism, which can generate faithful results compared to the expected model behaviour with sufficient accuracy.

The experiments consist of running the proposed PMPP model 200 times 
(i.e., 100 for cost and 100 for grid loss minimization where feasible) 
on all 44 test cases available in PGLib-OPF. 
The experiments were conducted on a 12 core hyper-threaded 3.5GHz Xeon CPU 128GB RAM desktop with a timeout period of 10 minutes for each run. The perturbation parameters are $\alpha=0.01$, $\beta=0.5$, $\e=1$ and $\lambda=30$. The above experiments were repeated by increasing $\alpha$ to 0.1 while keeping the rest unchanged.

\subsection{Solvability of PMPP}
\begin{figure}
\centerline{\includegraphics[scale=0.08]{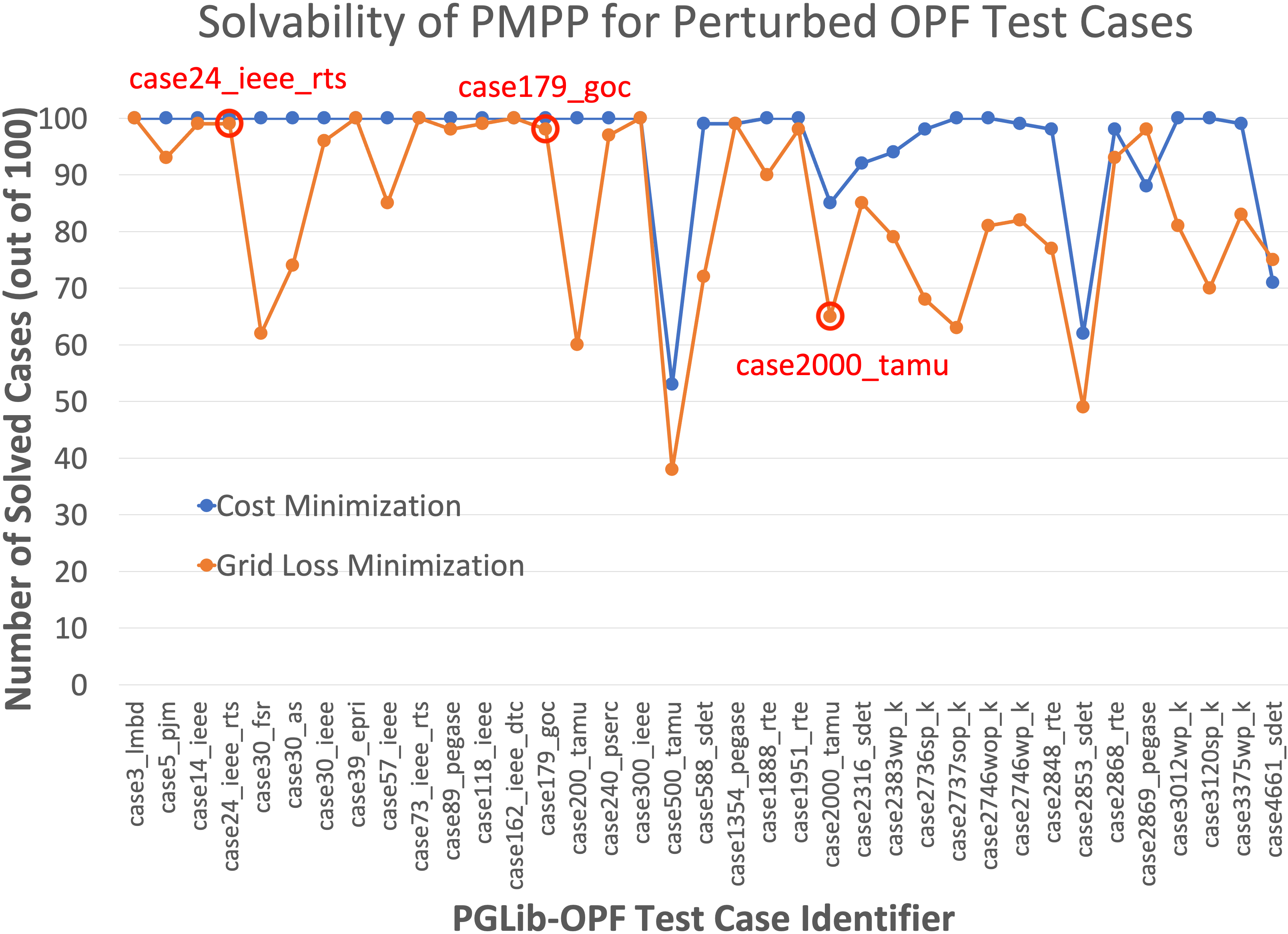}}
\caption{Results for PMPP solving PGLib-OPF test cases with differential privacy, where $\alpha=0.01$; $\e=1$; $\beta=0.5$; $\lambda=30$}
\label{Experiment}
\end{figure}

Fig.~\ref{Experiment} shows the results for $\alpha=0.01$ where a majority of test cases completed successfully up to the test case4661\_sdet and within a 6-day period. These are substantial results, as according to Model 1 and Algorithm 1, a nonlinear optimization problem with cubic polynomials is implied, which in a general case may not necessarily be solvable. Nevertheless, we successfully ran the mechanism for cases up to 4700 buses, a much larger number of buses than \cite{fioretto2020}, where only up to 118-buses were evaluated. 
For larger test cases such as case6468\_rte and beyond, 
we frequently encountered insolvable or timeout issues, which are not recorded in Fig.~\ref{Experiment}.

\begin{remark}
Overall, 
we observe that the results of cost and grid loss minimization are highly correlated, 
with the latter generating slightly worse results. 
It is worth noting that for test cases with a size of 4661 and below, 
we were able to solve optimal power flow test cases with a high probability. 
In more detail, 
96\% and 84\% of cases were successful for cost minimization and for grid loss minimization, respectively. 
\end{remark}

\subsection{Investigation of Three Test Cases}
\begin{figure*}
\subfloat{\includegraphics[scale=0.385, trim={0 0 0 1.4cm},clip]{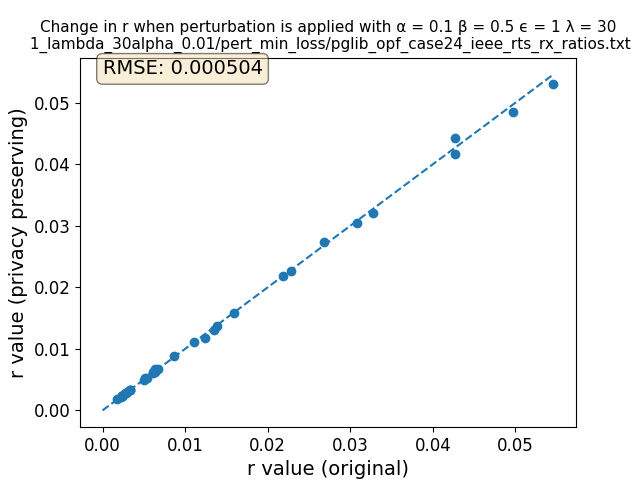}}
\hfill
\subfloat{\includegraphics[scale=0.385, trim={0 0 0 1.4cm},clip]{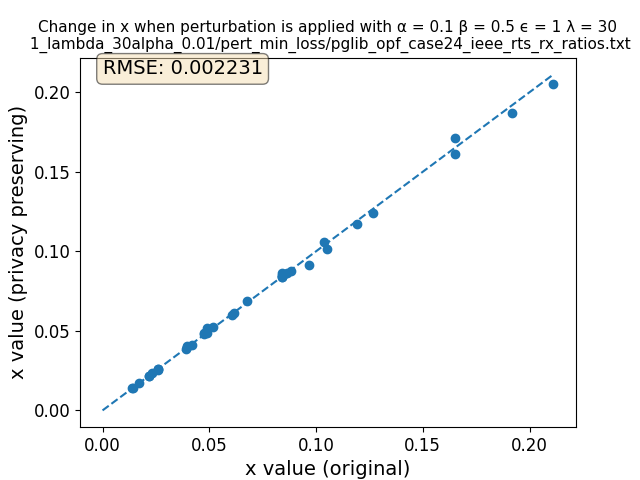}}
\subfloat{\includegraphics[scale=0.385, trim={0 0 0 1.4cm},clip]{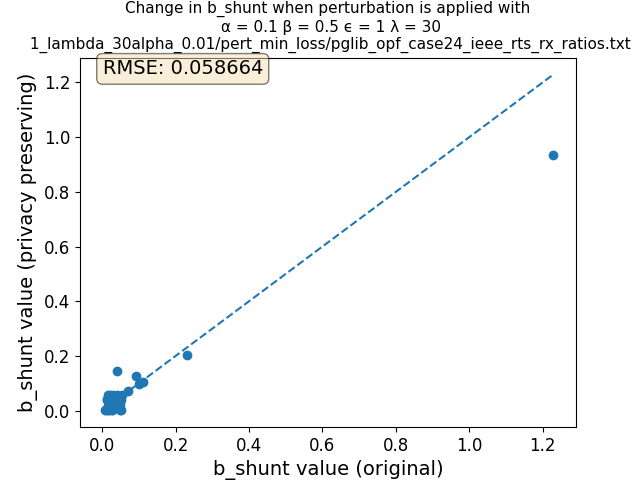}}
\caption{Utility of series resistance $r$, reactance $x$, and shunt susceptance (p.u.) perturbed values in 24-bus test case for $\alpha=0.01$}
\label{Experiment73}
\end{figure*}

To elaborate on the above results of solvability, 
we investigate three test cases from Fig.~\ref{Experiment}, 
where the percentage of the solved cases varied between 65\% and 99\% for the 200 runs in our experiments. In this Section all values are the per-unit (p.u.) values.
In Fig.~\ref{Experiment73}, 
we show\footnote{$z = Impedance = r+jx = 1/Admittance  = 1/y = 1/(g+jb)$} series resistance, $r$, series reactance, $x$, and the shunt susceptance $b^{sh}$ (as a reminder, shunt conductance is zero), respectively for a 24-bus test case, 
with privacy-preserved values versus original values with indistinguishability $\alpha=0.01$.
The deviation of the numerical results from the dashed diagonal line indicates the error, 
and hence the amount of accuracy of the privacy-preserved grid loss optimization. 
With a very small amount of perturbation in most $r$, $x$ and $b^{sh}$ values , 
this test case can produce valid solutions while maintaining the model constraints. Here, we also note the root-mean-square-error (RMSE) is respectively 0.000504 for resistance $r$, 0.002231 for reactance $x$, and 0.058664 for shunt susceptance $b^{sh}$.

\begin{figure*}
\subfloat{\includegraphics[scale=0.385, trim={0 0 0 1.4cm},clip]{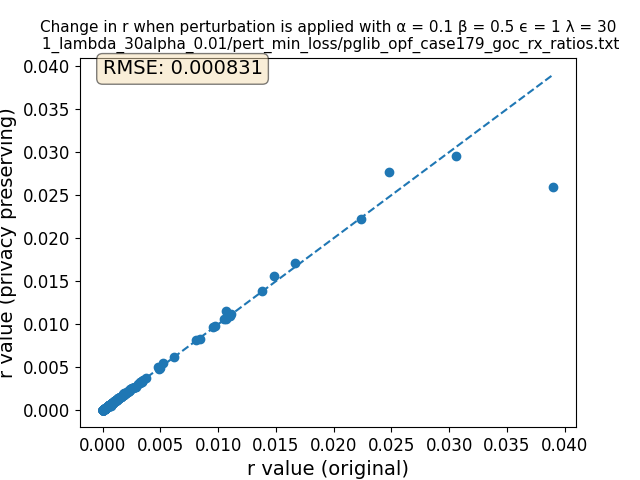}}
\hfill
\subfloat{\includegraphics[scale=0.385, trim={0 0 0 1.4cm},clip]{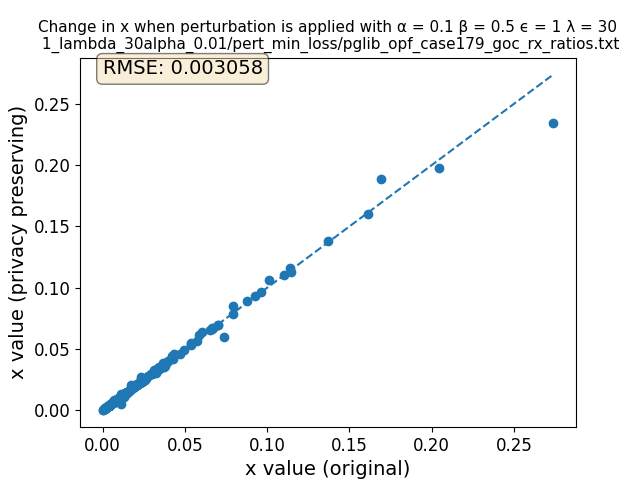}}
\subfloat{\includegraphics[scale=0.385, trim={0 0 0 1.4cm},clip]{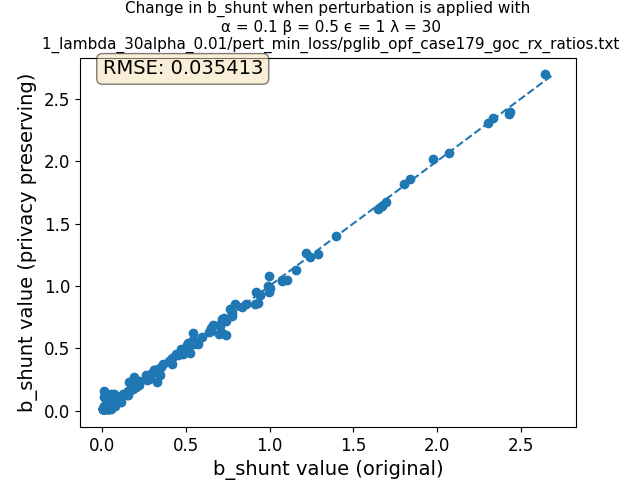}}
\caption{Utility of series resistance, reactance and shunt susceptance (p.u.) perturbed values in 179-bus test case for $\alpha=0.01$}
\label{Experiment162}
\end{figure*}

For a 179-bus test case demonstrated in Fig.~\ref{Experiment162},
we can see that a similar level of noise is required to be added to $r$, $x$ and $b^{sh}$  values, to reach the OPF objectives as per 24-buses. This results in a slightly higher RMSE than the 24-bus test case for series impedance, 0.000831 for $r$ and 0.003058 for $x$, with a slightly lower RMSE for shunt susceptance of 0.035413.

\begin{figure*}
\subfloat{\includegraphics[scale=0.385, trim={0 0 0 1.4cm},clip]{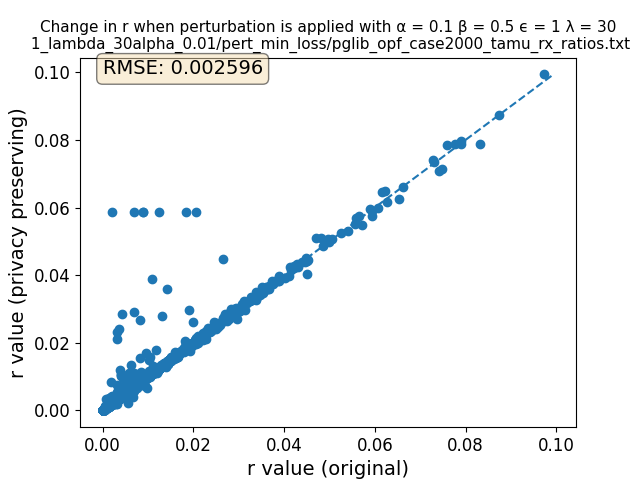}}
\hfill
\subfloat{\includegraphics[scale=0.385, trim={0 0 0 1.4cm},clip]{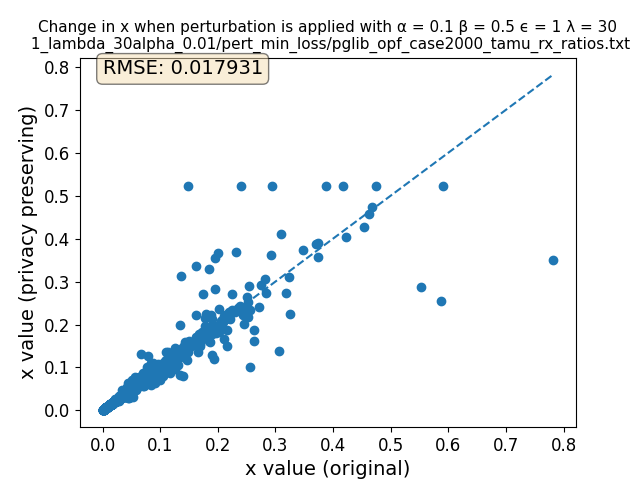}}
\subfloat{\includegraphics[scale=0.385, trim={0 0 0 1.4cm},clip]{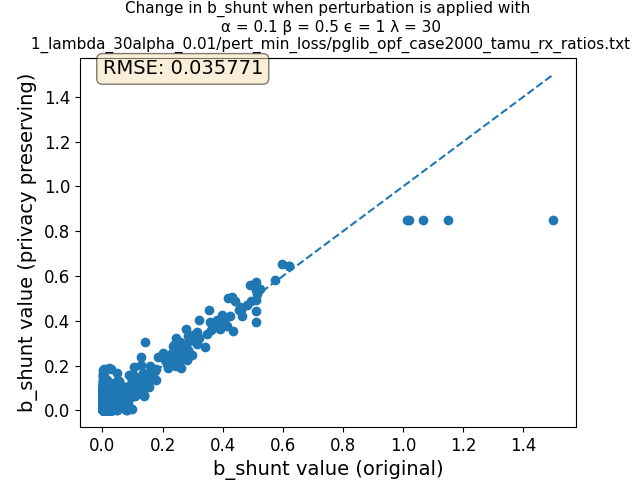}}
\caption{Utility of series resistance, reactance and shunt susceptance (p.u.) perturbed values in 2000-bus test case for $\alpha=0.01$}
\label{Experiment1354}
\end{figure*}

For an even larger test case of 2000 buses exhibited in Fig.~\ref{Experiment1354},
it can be seen that a larger amount of noise is needed for a majority of the $r$ and $x$ values such that their privacy can be protected with high utility while satisfying the overall optimization constraints. However the results are still substantially accurate with RMSE of 0.002596 and 0.017931 for series impedance, $r$ and $x$, and RMSE of 0.035771 for shunt susceptance $b^{sh}$.

\begin{figure*}
\subfloat{\includegraphics[scale=0.385, trim={0 0 0 1.4cm},clip]{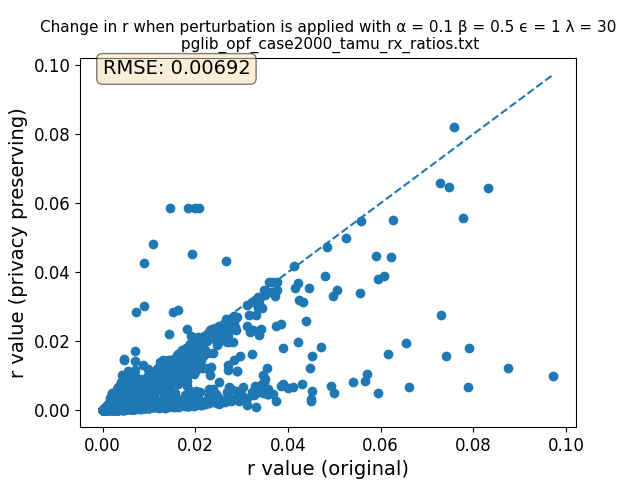}}
\hfill
\subfloat{\includegraphics[scale=0.385, trim={0 0 0 1.4cm},clip]{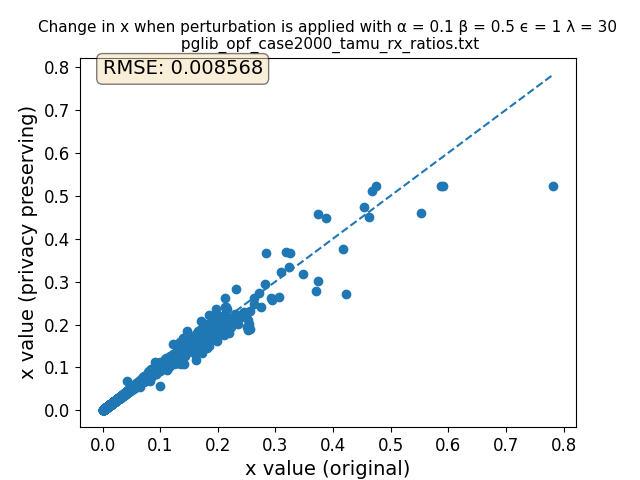}}
\subfloat{\includegraphics[scale=0.385, trim={0 0 0 1.4cm},clip]{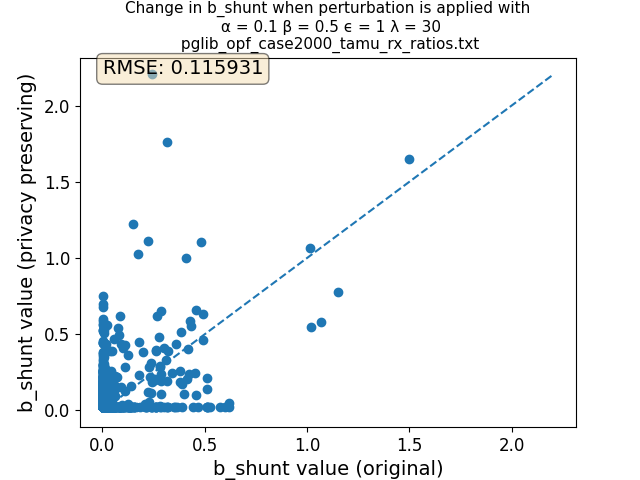}}
\caption{Utility of series resistance, reactance and shunt susceptance (p.u.) perturbed values in 2000-bus test case for $\alpha=0.1$}
\label{Experiment1354_2}
\end{figure*}

Depending on the data custodian's risk appetite, 
one may want to raise the level of perturbation, and further reduce risk by increasing the indistinguishability value ($\alpha$).  By increasing $\alpha$ from 0.01 to 0.1, 
we show the utility results for the test case of 2000 buses in Fig.~\ref{Experiment1354_2}, 
and we can observe both higher and lower level of perturbation of series $r$ and $x$ values compared with that displayed in~Fig.~\ref{Experiment1354}. 
Nevertheless, 
it still satisfies the model constraints. The RMSE is 0.00692 for series $r$ and 0.008568 for series $x$. The shunt susceptance has a reasonable RMSE of 0.115931, with the most values being close to $b^{sh}=0$, with some larger perturbation of true $b^{sh}=0$ values for $b^{sh}$.

\begin{remark}
We note that in all test cases illustrated in Figs. 2 -- 5, as the scale of series susceptance decreases in Algorithm 1, constraint ($s_5$) is reduced to a smaller scale according to $\mu_b$ and Algorithm 1, line 5, thus the magnitude of reactance $x$, and its perturbed noisy value, is larger and has more variability. This did not appear to affect the ability to solve the system.
\end{remark}

\section{Concluding Remarks}
We presented and validated a novel method for realistic differentially-private high voltage power flow data release, that is faithful to network physics, while protecting the network impedance data in a smart grid from being directly identified and still providing utility to any data analyst. We significantly extended and built on prior work to reduce the leakage of private information, better represented the transmission-side (in-terms of branch shunt susceptance and series reactance) and also scaled use case solvability to a significantly larger number of  buses (by over a factor of 10). Thus, we provided a feasible solution for transmission network providers to release privacy-preserving practical versions of commercially sensitive grid data, which we implemented and released in a software package on top of PowerModels.jl. Further work is to develop and implement a similar method for medium and low voltage distribution networks, where impedance is a matrix variable due to phase unbalance and electromagnetic coupling between phases. 

\bibliographystyle{IEEEtran}
\balance
\bibliography{privacy_energy_tech_TSG_ref}

\begin{thebibliography}{10}
\providecommand{\url}[1]{#1}
\csname url@samestyle\endcsname
\providecommand{\newblock}{\relax}
\providecommand{\bibinfo}[2]{#2}
\providecommand{\BIBentrySTDinterwordspacing}{\spaceskip=0pt\relax}
\providecommand{\BIBentryALTinterwordstretchfactor}{4}
\providecommand{\BIBentryALTinterwordspacing}{\spaceskip=\fontdimen2\font plus
\BIBentryALTinterwordstretchfactor\fontdimen3\font minus
  \fontdimen4\font\relax}
\providecommand{\BIBforeignlanguage}[2]{{%
\expandafter\ifx\csname l@#1\endcsname\relax
\typeout{** WARNING: IEEEtran.bst: No hyphenation pattern has been}%
\typeout{** loaded for the language `#1'. Using the pattern for}%
\typeout{** the default language instead.}%
\else
\language=\csname l@#1\endcsname
\fi
#2}}
\providecommand{\BIBdecl}{\relax}
\BIBdecl

\bibitem{fioretto2020}
F.~{Fioretto}, T.~W.~K. {Mak}, and P.~{Van Hentenryck}, ``Differential privacy
  for power grid obfuscation,'' \emph{IEEE Transactions on Smart Grid},
  vol.~11, no.~2, pp. 1356--1366, 2020.

\bibitem{dwork2014}
C.~Dwork and A.~Roth, ``The algorithmic foundations of differential privacy.''
  \emph{Foundations and Trends in Theoretical Computer Science}, vol.~9, no.
  3-4, pp. 211--407, 2014.

\bibitem{dwork2006}
C.~Dwork, F.~McSherry, K.~Nissim, and A.~Smith, ``Calibrating noise to
  sensitivity in private data analysis,'' \emph{Theory of cryptography
  conference}, pp. 265--284, 2006.

\bibitem{wood2013}
A.~J. Wood, B.~F. Wollenberg, and G.~B. Shebl{\'e}, \emph{Power generation,
  operation, and control}.\hskip 1em plus 0.5em minus 0.4em\relax John Wiley \&
  Sons, 2013.

\bibitem{dhople2015}
S.~V. Dhople, S.~S. Guggilam, and Y.~C. Chen, ``Linear approximations to ac
  power flow in rectangular coordinates,'' in \emph{2015 53rd Annual Allerton
  Conference on Communication, Control, and Computing (Allerton)}.\hskip 1em
  plus 0.5em minus 0.4em\relax IEEE, 2015, pp. 211--217.

\bibitem{purchala2005}
K.~Purchala, L.~Meeus, D.~Van~Dommelen, and R.~Belmans, ``Usefulness of dc
  power flow for active power flow analysis,'' in \emph{IEEE Power Engineering
  Society General Meeting, 2005}.\hskip 1em plus 0.5em minus 0.4em\relax IEEE,
  2005, pp. 454--459.

\bibitem{nationalgrideso}
\BIBentryALTinterwordspacing
``Monthly transmission loss data (4l),'' \emph{nationalgridESO}. [Online].
  Available:
  \url{https://www.nationalgrideso.com/balancing-data/monthly-transmission-loss-data-4l}
\BIBentrySTDinterwordspacing

\bibitem{aemo}
\BIBentryALTinterwordspacing
``Loss factors,'' \emph{Australian Energy Market Operator}. [Online].
  Available:
  \url{https://aemo.com.au/en/energy-systems/electricity/wholesale-electricity-market-wem/data-wem/loss-factors}
\BIBentrySTDinterwordspacing

\bibitem{powermodels}
C.~Coffrin, R.~Bent, K.~Sundar, Y.~Ng, and M.~Lubin, ``Powermodels.jl: An
  open-source framework for exploring power flow formulations,'' in \emph{2018
  Power Systems Computation Conference (PSCC)}, June 2018, pp. 1--8.

\bibitem{babaeine2019}
S.~Babaeinejadsarookolaee \emph{et~al.}, ``The power grid library for
  benchmarking ac optimal power flow algorithms,'' \emph{arXiv preprint
  arXiv:1908.02788 [math.OC]}, 2019.

\bibitem{coffrin2014n}
C.~Coffrin, D.~Gordon, and P.~Scott, ``{NESTA}, the {NICTA} energy system test
  case archive,'' \emph{arXiv preprint arXiv:1411.0359}, 2014.

\bibitem{bai2008}
\BIBentryALTinterwordspacing
``Semidefinite programming for optimal power flow problems,''
  \emph{International Journal of Electrical Power \& Energy Systems}, vol.~30,
  no.~6, pp. 383 -- 392, 2008. [Online]. Available:
  \url{http://www.sciencedirect.com/science/article/pii/S0142061507001378}
\BIBentrySTDinterwordspacing

\bibitem{hijazi2017c}
H.~Hijazi, C.~Coffrin, and P.~Van~Hentenryck, ``Convex quadratic relaxations
  for mixed-integer nonlinear programs in power systems,'' \emph{Mathematical
  Programming Computation}, vol.~9, no.~3, pp. 321--367, 2017.

\bibitem{jabr2006r}
R.~A. Jabr, ``Radial distribution load flow using conic programming,''
  \emph{IEEE transactions on power systems}, vol.~21, no.~3, pp. 1458--1459,
  2006.

\bibitem{coffrin2015qc}
C.~Coffrin, H.~L. Hijazi, and P.~Van~Hentenryck, ``The {QC} relaxation: A
  theoretical and computational study on optimal power flow,'' \emph{IEEE
  Transactions on Power Systems}, vol.~31, no.~4, pp. 3008--3018, 2015.

\bibitem{fioretto2018}
F.~Fioretto and P.~V. Hentenryck, ``Constrained-based differential privacy:
  Releasing optimal power flow benchmarks privately,'' in \emph{International
  Conference on the Integration of Constraint Programming, Artificial
  Intelligence, and Operations Research}.\hskip 1em plus 0.5em minus
  0.4em\relax Springer, 2018, pp. 215--231.

\bibitem{mak2020}
T.~W.~K. {Mak}, F.~{Fioretto}, L.~{Shi}, and P.~{Van Hentenryck},
  ``Privacy-preserving power system obfuscation: A bilevel optimization
  approach,'' \emph{IEEE Transactions on Power Systems}, vol.~35, no.~2, pp.
  1627--1637, 2020.

\bibitem{dvorkin2020b}
V.~Dvorkin, F.~Fioretto, P.~V. Hentenryck, J.~Kazempour, and P.~Pinson,
  ``Differentially private convex optimization with feasibility guarantees,''
  \emph{arXiv preprint arXiv:2006.12338 [cs.CR]}, 2020.

\bibitem{dvorkin2020a}
V.~Dvorkin, F.~Fioretto, P.~Van~Hentenryck, P.~Pinson, and J.~Kazempour,
  ``Differentially private optimal power flow for distribution grids,''
  \emph{IEEE Transactions on Power Systems}, 2020.

\bibitem{zhou2019d}
F.~Zhou, J.~Anderson, and S.~H. Low, ``Differential privacy of aggregated {DC}
  optimal power flow data,'' in \emph{2019 American Control Conference
  (ACC)}.\hskip 1em plus 0.5em minus 0.4em\relax IEEE, 2019, pp. 1307--1314.

\bibitem{bienstock2019}
D.~{Bienstock} and A.~{Shukla}, ``Variance-aware optimal power flow: Addressing
  the tradeoff between cost, security, and variability,'' \emph{IEEE
  Transactions on Control of Network Systems}, vol.~6, no.~3, pp. 1185--1196,
  2019.

\bibitem{ryu2021}
M.~Ryu and K.~Kim, ``A privacy-preserving distributed control of optimal power
  flow,'' \emph{arXiv preprint arXiv:2102.02276}, 2021.

\bibitem{karapetyan2017a}
A.~Karapetyan, S.~K. Azman, and Z.~Aung, ``Assessing the privacy cost in
  centralized event-based demand response for microgrids,'' in \emph{2017 IEEE
  Trustcom/BigDataSE/ICESS}.\hskip 1em plus 0.5em minus 0.4em\relax IEEE, 2017,
  pp. 494--501.

\bibitem{zhao2014}
J.~{Zhao}, T.~{Jung}, Y.~{Wang}, and X.~{Li}, ``Achieving differential privacy
  of data disclosure in the smart grid,'' in \emph{IEEE INFOCOM}, 2014, pp.
  504--512.

\bibitem{eibl2017d}
G.~Eibl and D.~Engel, ``Differential privacy for real smart metering data,''
  \emph{Computer Science-Research and Development}, vol.~32, no. 1-2, pp.
  173--182, 2017.

\bibitem{liao2019}
X.~{Liao}, P.~{Srinivasan}, D.~{Formby}, and R.~A. {Beyah}, ``{Di}-{PriDA}:
  Differentially private distributed load balancing control for the smart
  grid,'' \emph{IEEE Transactions on Dependable and Secure Computing}, vol.~16,
  no.~6, pp. 1026--1039, 2019.

\bibitem{smith2020}
D.~{Smith} \emph{et~al.}, ``Privacy-preserved optimal energy trading,
  statistics, and forecasting for a neighborhood area network,'' \emph{IEEE
  Computer Magz.}, vol.~53, no.~5, pp. 25--34, 2020.

\bibitem{han2016}
S.~{Han}, U.~{Topcu}, and G.~J. {Pappas}, ``Differentially private distributed
  constrained optimization,'' \emph{IEEE Transactions on Automatic Control},
  vol.~62, no.~1, pp. 50--64, 2017.

\bibitem{arpa-e}
\BIBentryALTinterwordspacing
``Data repository for power system models,'' \emph{ARPA-E}. [Online].
  Available:
  \url{https://arpa-e.energy.gov/technologies/projects/data-repository-power-system-models}
\BIBentrySTDinterwordspacing

\bibitem{nemweb}
\BIBentryALTinterwordspacing
``Market data nemweb,'' \emph{Australian Energy Market Operator}. [Online].
  Available:
  \url{https://www.aemo.com.au/energy-systems/electricity/national-electricity-market-nem/data-nem/market-data-nemweb}
\BIBentrySTDinterwordspacing

\bibitem{ga-substation}
\BIBentryALTinterwordspacing
``Electricity transmission substations,'' \emph{Geoscience Australia}.
  [Online]. Available: \url{http://pid.geoscience.gov.au/dataset/ga/83173}
\BIBentrySTDinterwordspacing

\bibitem{ga-line}
\BIBentryALTinterwordspacing
``Electricity transmission lines,'' \emph{Geoscience Australia}. [Online].
  Available: \url{http://pid.geoscience.gov.au/dataset/ga/83105}
\BIBentrySTDinterwordspacing

\bibitem{bienstock2010nk}
D.~Bienstock and A.~Verma, ``The {N}-k problem in power grids: New models,
  formulations, and numerical experiments,'' \emph{SIAM Journal on
  Optimization}, vol.~20, no.~5, pp. 2352--2380, 2010.

\bibitem{salmeron2009worst}
J.~Salmeron, K.~Wood, and R.~Baldick, ``Worst-case interdiction analysis of
  large-scale electric power grids,'' \emph{IEEE Transactions on power
  systems}, vol.~24, no.~1, pp. 96--104, 2009.

\bibitem{liu2011false}
Y.~Liu, P.~Ning, and M.~Reiter, ``False data injection attacks against state
  estimation in electric power grids,'' \emph{ACM Transactions on Information
  and System Security (TISSEC)}, vol.~14, no.~1, pp. 1--33, 2011.

\bibitem{yuan2011modeling}
Y.~Yuan, Z.~Li, and K.~Ren, ``Modeling load redistribution attacks in power
  systems,'' \emph{IEEE Transactions on Smart Grid}, vol.~2, no.~2, pp.
  382--390, 2011.

\bibitem{gorman2009e}
\BIBentryALTinterwordspacing
``Electricity {Grid} in {U.S.} penetrated by spies,'' \emph{The Wall Street
  Journal}. [Online]. Available: \url{https://www.wsj.com/articles/
  SB123914805204099085}
\BIBentrySTDinterwordspacing

\bibitem{zetter2016i}
\BIBentryALTinterwordspacing
``Inside the cunning, unprecedented hack of {U}kraine's power grid,''
  \emph{Wired Magazine}. [Online]. Available:
  \url{https://www.wired.com/2016/03/inside-cunning-unprecedented-hack-ukraines-power-grid/}
\BIBentrySTDinterwordspacing

\bibitem{mehta2016}
D.~Mehta, D.~K. Molzahn, and K.~Turitsyn, ``Recent advances in computational
  methods for the power flow equations,'' in \emph{2016 American Control
  Conference (ACC)}.\hskip 1em plus 0.5em minus 0.4em\relax IEEE, 2016, pp.
  1753--1765.

\end{thebibliography}


\begin{thebibliography}{10}

\csname url@samestyle\endcsname
\providecommand{\newblock}{\relax}
\providecommand{\bibinfo}[2]{#2}
\providecommand{\BIBentrySTDinterwordspacing}{\spaceskip=0pt\relax}
\providecommand{\BIBentryALTinterwordstretchfactor}{4}
\providecommand{\BIBentryALTinterwordspacing}{\spaceskip=\fontdimen2\font plus
\BIBentryALTinterwordstretchfactor\fontdimen3\font minus
  \fontdimen4\font\relax}
\providecommand{\BIBforeignlanguage}[2]{{%
\expandafter\ifx\csname l@#1\endcsname\relax
\typeout{** WARNING: IEEEtran.bst: No hyphenation pattern has been}%
\typeout{** loaded for the language `#1'. Using the pattern for}%
\typeout{** the default language instead.}%
\else
\language=\csname l@#1\endcsname
\fi
#2}}
\providecommand{\BIBdecl}{\relax}
\BIBdecl





\end{thebibliography}

\end{document}